\documentclass[useAMS,usenatbib]{mn2e}

\usepackage{epsfig}

\usepackage{times}

\title[The slope of the GRB Variability/Peak Luminosity Correlation]
{The slope of the GRB Variability/Peak Luminosity Correlation}
\author[C. Guidorzi et al.]{C. Guidorzi$^{1,2}$\thanks{E-mail:
crg@astro.livjm.ac.uk},
F. Frontera$^{2,3}$, E. Montanari$^{2,4}$, F. Rossi$^{2}$, L. Amati$^{3}$,
\newauthor A. Gomboc$^{5}$, C.G. Mundell$^{1}$\\
$^{1}$Astrophysics Research Institute, Liverpool John Moores University,
Twelve Quays House, Birkenhead, CH41 1LD, UK\\
$^{2}$Dipartimento di Fisica, Universit\`a di Ferrara, via Saragat 1,
44100 Ferrara, Italy\\
$^{3}$Istituto di Astrofisica Spaziale e Fisica Cosmica of Bologna, INAF,
Via Gobetti 101, 40129 Bologna, Italy\\
$^{4}$ITA ``Calvi'', Finale Emilia (MO), Italy\\
$^{5}$Faculty of Mathematics and Physics, University in Ljubljana,
Jadranska 19, 1000 Ljubljana, Slovenia}
\begin{document}

\date{}

\pagerange{\pageref{firstpage}--\pageref{lastpage}} \pubyear{2002}

\maketitle

\label{firstpage}

\begin{abstract}
First using a sample of 32 GRBs with known redshift (Guidorzi et al. 2005)
and then a sample of 551 BATSE GRBs with derived pseudo-redshift (Guidorzi 2005),
the time variability/peak luminosity correlation ($V$ vs. $L$), originally 
found by Reichart et al. (2001) using a sample of 18 GRBs, was tested. For both 
samples the correlation is still found but less relevant due to a much 
higher spread of the data.
Assuming a straight line in the $\log{L}$--$\log{V}$ plane ($\log{L} = m \log{V} + b$),
as done by Reichart et al., both Guidorzi et al. and Guidorzi found that
the line slope for both samples is much lower than that derived by
Reichart et al.: $m = 1.3_{-0.4}^{+0.8}$ (Guidorzi et al. 2005),
$m = 0.85\pm 0.02$ (Guidorzi 2005), $m = 3.3^{+1.1}_{-0.9}$ (Reichart et al. 2001). 
Reichart \& Nysewander (2005) discuss our results and attribute the different
slope to the fact we do not take into account in the fit the variance of the
sample (also called slop), and demonstrate that, using the method presented
by Reichart (2001), the expanded data set of Guidorzi et al. (2005)
in $\log{L}$--$\log{V}$ plane is still well described by a line with slope 
$m = 3.4^{+0.9}_{-0.6}$.
Here we compare the results of two methods accounting for the slop of the sample,
the method implemented by Reichart (2001) and that by D'Agostini (2005).
We demonstrate that the method used by Reichart et al. (2001) to estimate
the straight line slope, provides an inconsistent estimate of the parameter
when the sample variance is comparable with the interval of values 
covered by the GRB variability. We also demonstrate that, using the D'Agostini
method, the slope of the $\log{L}$--$\log{V}$ 
correlation is still consistent with that derived by us earlier 
and inconsistent with that derived by Reichart \& Nysewander (2005). 
Finally we discuss the implications on the interpretations
proposed for the $V-L$ correlation and show that our results are in agreement
with the peak energy/variability correlation found by Lloyd-Ronning \& Ramirez-Ruiz 
(2002) and the peak energy/peak luminosity correlation 
(Yonetoku et al. 2004; Ghirlanda et al. 2005). 
\end{abstract}


\section{Introduction}
\label{s:intro}

Most of our knowledge about the Gamma-Ray Burst (GRB) phenomenon is derived 
from their spectra and light curve profiles, but it is recognised
that other observational probes (e.g., polarisation of the gamma-rays) would
give key  information to the solution of the GRB enigma. Among these probes,
it is recognised the importance of the erratic time variability of the 
GRB time profiles. For example, in the GRB internal shock model, 
not very variable radiation is expected to be produced at radii 
lower than the photosphere radius in which the shocks remain optically thin 
to pairs \citep{Kobayashi02}, while highly
variable radiation is expected to be produced in shocks above this radius
\citep{Meszaros02}. Also in the sub-jet model by \citet{Ioka01}, time
variability is expected and its amplitude related to the viewing angle of the burst.

A key objective of the study of GRBs is to
establish whether GRBs can be reliably used as standard candles and to
determine the optimal relationship between observed and intrinsic
properties.
The recent discovery of GRBs with bright afterglows at redshifts $z>6$
highlights their power as probes of the high redshift Universe
\citep{Haislip06,Kawai06}, but
spectroscopically-confirmed redshifts are only a fraction.
In contrast, the characteristics of gamma-ray light
curves, which are available for all GRBs, offer
a potentially independent estimate of luminosity distance for
statistically-significant samples.

An important problem, however, in deriving the intrinsic correlations
is the interpretation of scatter in correlations, which may be
produced by measurement methods, construction of samples with
properties measured by satellites with differing response functions,
small samples and different statistical analysis methods and 
intrinsic physical differences in the GRB population \citep{Nava06}.
As sample sizes slowly increase, addressing these issues
remains critical for the correct inference of intrinsic GRB properties
and thus their use as cosmological probes.

%Time variability studies of GRBs have been performed by various
%authors, e.g. \citet{Beloborodov00}.
In this paper, we concentrate on a long-standing empirical relation
that initially suggested a possible Cepheid-like correlation between
gamma-ray variability and peak luminosity of GRBs
\citep{Reichart01,Fenimore00}.

\citet{Reichart01} (hereafter R01), using a robust measure $V$ of the
GRB variability, for a sample of 13 GRBs with known redshift, found 
that in the GRB rest frame this measure  is correlated with the GRB
peak luminosity $L$.
In the $\log{L}$--$\log{V}$ plane the correlation was modelled with a linear
function $\log L = m \log V  + b$ with the slope of the straight line $m = 
3.3^{+1.1}_{-0.9}$ and a sample variance along $V$ of $\sigma_{\log{V}}=0.18$,
both parameters being obtained with the method described by
\citet{Reichart01b} (hereafter called Reichart method).
This method was proposed to fit
data sets affected by a sample variance in addition to the statistical
variance (called "intrinsic variance") of each data point.
Recently, first \citet{Guidorzi05a} (hereafter GFM05) and then \citet{Guidorzi05b}
(hereafter G05)
tested this correlation first using an extended sample of 32 GRBs with known redshift
(GFM05), and then with 551 GRBs detected by {\em CGRO}/BATSE \citep{Paciesas99}
for which a pseudo-redshift was derived by G05 exploiting the spectral lag-luminosity 
correlation \citep{Norris00, Norris02,Band04}. In both cases, the
correlation was confirmed and in the $\log V$--$\log L$ plane
the slope of the straight line was found much lower than that derived by R01 
($m = 1.3_{-0.4}^{+0.8}$ derived by GFM05; $m = 0.85\pm 0.02$ derived by G05).
It was also found that, with the sample variance neglected, the straight line did not 
provide a good description of the data ($\chi^2/{\rm dof} = 1167/30$ and
$\chi^2/{\rm dof} = 4238/549$ for the samples considered by GFM05 and G05,
respectively).
Neglecting the sample variance was correctly questioned by \citet{Reichart05} 
(hereafter RN05), who however show that, also using the extended data set of GFM05, 
the $m$ slope is given by $m = 3.4_{-0.6}^{+0.9}$, still in perfect agreement with
the original value found by R01 and in strong disagreement with
the value found by GFM05. They attribute this disagreement to the fact
that GFM05 do not include among the parameters of the fit the sample variance, which
they estimate to be $\sigma_{\log{V}}=0.20\pm0.04$. 
In an unrefereed note, \citet{Reichart05_comm} takes issue with the most recent
paper by G05, who confirms the results previously found by GFM05.

In this paper, we discuss the Reichart method and compare the results
with those obtained following the treatment by \citet{Dagostini05}
(hereafter called ``the D'Agostini method''), which deals with the same
problem of fitting data points affected by extrinsic scatter in addition
the intrinsic uncertainties along both axes.
We show with numerical simulations that the Reichart method
provides an inconsistent estimate of the $m$ slope in the specific case
of the $\log V$--$\log L$ data set, we discuss the likely
reason for this inconsistency. We show that our original results are 
substantially confirmed by the D'Agostini method even taking into account
the sample variance. Finally we discuss the subsequent implications
for the inferred physics of GRB
central engines and their relativistic outflows. The usefulness of the
so-called variability/peak luminosity correlation is discussed in the
broader context of recently discovered correlations between other
observed and derived properties of GRBs.

%%%%%%%%%%%%%%%%%%%%%%%%%%%%%%%%%%%%%%%%%%%%%%%%%%%%%%%%%
\section[]{The Reichart method}
\label{s:meth}
%%%%%%%%%%%%%%%%%%%%%%%%%%%%%%%%%%%%%%%%%%%%%%%%%%%%%%%%%
The Reichart method has as starting point the well known Bayes 
theorem that, for inference of physical parameters, is widely
discussed in several text books (see, e.g., D'Agostini 2003).
It  states that the conditional probability 
that a set of parameters \mbox{\boldmath$\theta$}$=\theta_1,\theta_2,...\theta_n$  
takes a particular value $\mbox{\boldmath$\theta$}^0$, given a data set $D$, 
whose values depend on \mbox{\boldmath$\theta$}
(for instance, N independent observations of a quantity $X$), is given by:
\begin{equation}
p(\mbox{\boldmath$\theta$} | D I) = \frac{p(\mbox{\boldmath$\theta$} | I) \ p(D | \mbox{\boldmath$\theta$} I)}{p(D | I)}
\end{equation}
where $I$ is the available prior information,
$p(\mbox{\boldmath$\theta$} |I)$ is the probability 
distribution of \mbox{\boldmath$\theta$} on the basis of the information $I$, 
and $p(D |\mbox{\boldmath$\theta$} I)$ is the conditional probability of getting the 
measured data set given the value $\mbox{\boldmath$\theta$}^0$ and the
information $I$. The probability distribution  $p(\mbox{\boldmath$\theta$} |I)$ is
called prior probability, while the probability distribution
$p(D |\mbox{\boldmath$\theta$} I)$  is called likelihood function.
The probability $p(D | I)$ is introduced as  normalization factor.
In the case that no prior information is known, $p(\mbox{\boldmath$\theta$}|I)$ is generally
assumed to be a uniform distribution and the range of the possible values of
\mbox{\boldmath$\theta$} are those logically allowed. In this case the posterior probability
and the likelihood function are equivalent.
 
Reichart (2001) concentrates on the derivation of the
prior probability $p(\mbox{\boldmath$\theta$}|I)$ in the special case in which for 
two of the $N$ parameters, $x$ and $y$ (in our specific case $x = \log V$ and 
$y = \log L$), we have a set of $N$ pairs of measurements, from which it appears that 
the two are correlated with
$y = y_c(x; \theta_m)$, were $\theta_m$, with $m = 1, 2$, \ldots, $M$, are M 
intermediate parameters that describe the curve $y_c(x;\theta_m)$ (in our case 
$y = m x + b$, $M = 2$). Correctly 
the scatter of the parameter values $x$ and $y$ around  the curve is assumed to 
be due partly to 
the measurement errors (intrinsic scatter) and partly due to weaker dependences of 
either parameters $x$ or $y$ on other, yet unmeasured, and even yet unknown, 
parameters (extrinsic scatter or sample scatter).
Both the scatters for $x$ and $y$, are assumed by Reichart to be normally distributed 
and uncorrelated, with unknown standard deviations $\sigma_x$ and $\sigma_y$
of the extrinsic scatter of $x$ and  $y$, respectively. 
The conditional probability of the  values of the parameters $\theta_m$, $\sigma_x$,
$\sigma_y$ given the measured set of data points and their uncertainties 
is thus derived under simplifying assumptions (see Section 2.2.2 of Reichart 2001), 
among which that the curve $y = y_c(x; \theta_m)$ can be approximated by a 
straight line ($y \approx y_{t,n} + s_{t,n}(x-x_{t,n})$). The result is the
following (eq.~43 of Reichart 2001):
\begin{eqnarray}
\lefteqn{ p(\theta_m, \sigma_x, \sigma_y | x_n, y_n, \sigma_{x,n}, \sigma_{y,n}) \approx \prod_{n=1}^N \sqrt{1+s_{t,n}^2} \ G_n \big[ y_n, y_{t,n} +{}} \nonumber\\
& & {}+s_{t,n}(x_n - x_{t,n}), \sqrt{\sigma_y^2 + \sigma_{y,n}^2 + s_{t,n}^2 (\sigma_x^2 + \sigma_{x,n}^2)} \big]
\label{e:reichart}
\end{eqnarray}
where $\sigma_{x,n}^2$ and $\sigma_{y,n}^2$ are the intrinsic variances of the $N$
pairs of data points $x_n$ and $y_n$ ($n = 1, 2$, \ldots, $N$), respectively;
$(x_{t,n}, y_{t,n})$ is the point on the curve $y=y_c(x; \theta_m)$ which
maximises the two-dimensional Gaussian $G_n(x, x_n, \sqrt{\sigma_x^2+\sigma_{x,n}^2})
\times G_n(y, y_n, \sqrt{\sigma_y^2+\sigma_{y,n}^2})$.

In order to simplify the derivation of the prior probability in this special case,
eq.~\ref{e:reichart} is assumed to be a likelihood function and the maximum 
likelihood method is applied to constrain the values of the intermediate parameters
$\theta_m$, $\sigma_x$ and $\sigma_y$ and the uncertainty in the values of $\theta_m$.

%%%%%%%%%%%%%%%%%%%%%%%%%%%%%%%%%%%%%%%%%%%%%%%%%%%%%%%%%
\subsection{Application of the Reichart method to the
Luminosity--Variability correlation}
\label{s:appl}
%%%%%%%%%%%%%%%%%%%%%%%%%%%%%%%%%%%%%%%%%%%%%%%%%%%%%%%%%

We now apply the Reichart method in the specific case of the 
variability--luminosity correlation test, showing that we are capable
of reproducing the results obtained by R01 and RN05. In the case
of this test, $N$ pairs of measured values  of 
$V_n$ and $L_n$ (one per each burst) are available, and, in 
the eq.~\ref{e:reichart}, $x_n=\log{(V_n/\bar{V})}$, 
$y_n=\log{(L_n/\bar{L})}$ ($\bar{V}$ and $\bar{L}$ being the correspondent 
median values), $\sigma_{x,n} = \sigma_{\log{V_n}}$ and 
$\sigma_{y,n}=\sigma_{\log{L_n}}$.
Having modelled the correlation by a straight line:
\begin{equation}
\displaystyle y(x) \ = \ m\,x \ + \ q
\label{eq:PL}
\end{equation}
from eq.~\ref{e:reichart}, it is possible to derive the log--likelihood function:
\begin{eqnarray}
\displaystyle \log{p(m,q,\sigma_x,\sigma_y|\{x_i,y_i,\sigma_{x,i},\sigma_{y,i}\})}
=~~~~~~~~~~~~~~~~~~~~~~~\nonumber\\
~~~~\frac{1}{2}\,\sum_{i=1}^N \Big[\log{\Big(\frac{1+m^2}{2 \pi (\sigma_y^2 + m^2\,\sigma_{x}^2 + \sigma_{y,i}^2 + m^2\,\sigma_{x,i}^2)}\Big)}\quad +\nonumber\\
~~~~-\quad\frac{(y_i - m\,x_i - q)^2}{\sigma_y^2 + m^2\,\sigma_{x}^2 + \sigma_{y,i}^2 + m^2\,\sigma_{x,i}^2}\Big]
\label{eq:prior}
\end{eqnarray}
in which, without loss of generality, it is possible to assume either
$\sigma_x=0$, or, alternatively, $\sigma_y=0$ (both appear only in the term 
$\sigma_y^2+m^2\,\sigma_x^2$). 

Assuming $\sigma_x=0$ in eq.~\ref{eq:prior}, with the maximum likelihood
method the free parameters $m$, $q$, and $\sigma_y$ can be derived.
Alternatively, if one assumes $\sigma_y=0$, the value for $\sigma_x$ 
can then be determined from $\sigma_x=\sigma_y/m$, where $\sigma_y$ is the
value obtained in the previous case under the assumption of null $\sigma_x$.

\begin{figure}
\includegraphics[width=8.5cm]{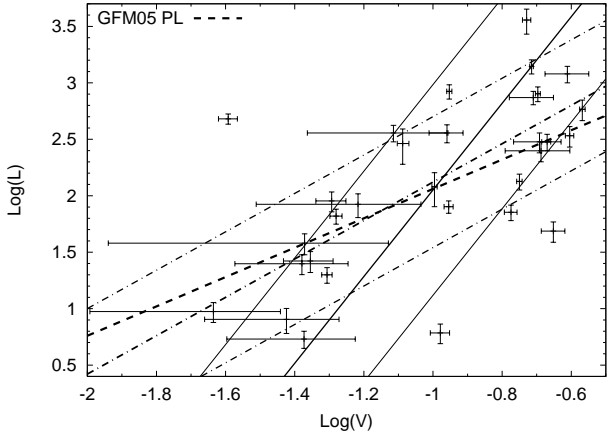}
\caption{Variability vs. Peak Luminosity for the 32 GRBs with
known redshift considered by GFM05. Also shown are the best fit  curves
and 1$\sigma$ regions. {\it Solid lines}: best fit results with
the Reichart method. {\it Dashed lines}: results obtained by GFM05.
{\it Dashed-dotted lines}: best fit results with the D'Agostini method.}
\label{f:z}
\end{figure}

\begin{figure}
\includegraphics[width=8.5cm]{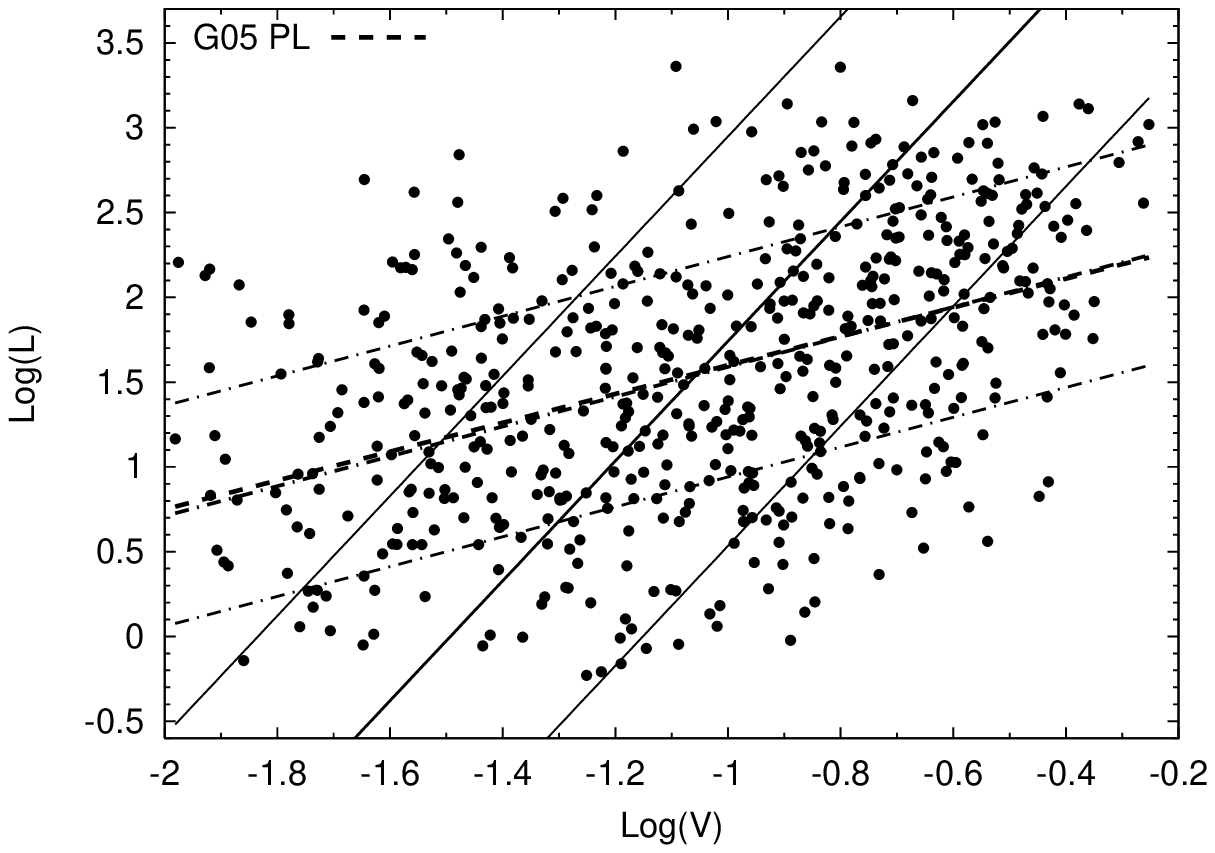}
\caption{Variability vs. Peak Luminosity for the 551 BATSE GRBs with
pseudo-redshift considered by G05. Also shown are the best fit  curves
and 1$\sigma$ regions. {\it Solid lines}: best fit results with
the Reichart method. {\it Dashed lines}: results obtained by G05.
{\it Dashed-dotted lines}: best fit results with the D'Agostini method.
}
\label{f:pz}
\end{figure}

For our extended sample of 32 GRBs with known redshift (GFM05), we find that 
at 90\% confidence level, the best-fitting parameter values
are the following: $m= 3.8_{-1.1}^{+2.8}$, $q=0.07_{-0.33}^{+0.32}$, 
$\sigma_y = \sigma_{\log L} = 0.93_{-0.29}^{+0.77}$ (alternatively, 
$\sigma_x = \sigma_{\log V} \approx 0.24$) (solid lines in Fig.~\ref{f:z}).
As can be seen, all these values
are in excellent agreement with those derived by RN05 (they report the 1$\sigma$ 
uncertainty). Similarly, for the sample of 551 BATSE GRBs with
pseudo-redshift (G05), we find $m=3.5_{-0.4}^{+0.6}$, 
$q=0.15_{-0.9}^{+0.9}$, $\sigma_y=1.21_{-0.15}^{+0.19}$,
or $\sigma_x \approx 0.35$ (solid lines in Fig.~\ref{f:pz}).
All these results are also summarised
in Table~\ref{tab:fit_results}.
%
%Figure 1
%
%
%
%Figure 2
%
%
%
% Table 1
%
\begin{table*}
\centering
  \caption{Best-fitting parameters obtained with different methods for the
GRB samples used by GFM05 and G05. The confidence intervals are at 90\%.}
  \label{tab:fit_results}
  \begin{tabular}{llllll}
\hline
GRB Set       & Method & $m$ & $q$ & $\sigma_y$     & $\sigma_x$    \\ 
              &        &     &     & ($\sigma_x=0$) & ($\sigma_y=0$)\\
\hline
 32 (from GFM05) & Reichart   & $3.8_{-1.1}^{+2.8}$ & $0.07_{-0.33}^{+0.32}$ & $0.93_{-0.29}^{+0.77}$
& $\sim0.24$\\
32 (from GFM05) & D'Agostini & $1.7_{-0.4}^{+0.4}$ & $0.07_{-0.19}^{+0.18}$ & $0.58_{-0.12}^{+0.15}$
& $\sim0.34$\\
32 (from GFM05) & {\tt fitexy}  & $1.9\pm0.1^{{\rm (a)}}$ & $0.14\pm0.02^{{\rm (a)}}$ & --
& --\\
551 (from G05)   & Reichart   & $3.5_{-0.4}^{+0.6}$ & $0.15_{-0.9}^{+0.9}$ & $1.21_{-0.15}^{+0.19}$
& $\sim0.35$\\
551 (from G05)   & D'Agostini & $0.88_{-0.13}^{+0.12}$ & $0.01_{-0.03}^{+0.03}$ & $0.65_{-0.04}^{+0.04}$
& $\sim0.74$\\
551 (from G05)   & {\tt fitexy}  & $1.37\pm0.02^{{\rm (a)}}$ & $0.09\pm0.01^{{\rm (a)}}$ & --
& --\\
\hline
\end{tabular}
\begin{list}{}{}
\item[$^{\rm (a)}$] $1\sigma$ confidence interval.
\end{list}
\end{table*}

%%%%%%%%%%%%%%%%%%%%%%%%%%%%%%%%%%%%%%%%%%%%%%%%%%%%%%%%%
\section[]{An unbiased method}
\label{s:dago_meth}
%%%%%%%%%%%%%%%%%%%%%%%%%%%%%%%%%%%%%%%%%%%%%%%%%%%%%%%%%
\citet{Dagostini05} addressed the same problem described in the
previous section, i.e. how to perform a linear fit between two
data sets with errors on both axes and with an extra variance.
Similarly to the Reichart method, the D'Agostini method
is based on the parametric inference typical of
the Bayesian approach. However, the resulting log-likelihood
derived by \citet{Dagostini05} (see eq. 35 and 52 therein) differs
from that by Reichart, reported in \ref{e:reichart}, for just one term:
in the D'Agostini likelihood function the term $(1+m^2)$ is just 1.
For a detailed description of the D'Agostini method and of its derivation
we address the reader to the original paper. We will demonstrate that, unlike
the Reichart method, the likelihood function derived by D'Agostini provides
unbiased estimates of the unknown parameters.

According to D'Agostini (2005), the log-likelihood function 
for the case here considered, is considered is given by
\begin{eqnarray}
\displaystyle \log{p(m,q,\sigma_x,\sigma_y|\{x_i,y_i,\sigma_{x,i},\sigma_{y,i}\})}
=~~~~~~~~~~~~~~~~~~~~~~~\nonumber\\
~~~~\frac{1}{2}\,\sum_{i=1}^N \Big[\log{\Big(\frac{1}{2 \pi (\sigma_y^2 + m^2\,\sigma_{x}^2 + \sigma_{y,i}^2 + m^2\,\sigma_{x,i}^2)}\Big)}\quad +\nonumber\\
~~~~-\quad\frac{(y_i - m\,x_i - q)^2}{\sigma_y^2 + m^2\,\sigma_{x}^2 + \sigma_{y,i}^2 + m^2\,\sigma_{x,i}^2}\Big]
\label{eq:prior_dago}
\end{eqnarray}
Using this equation, in the case of the 32 GRBs of GFM05, at 90\% confidence
level we find the results reported in Table~\ref{tab:fit_results}. 
In particular we find a slope $m= 1.7\pm0.4$ against a value 
$m= 3.8_{-1.1}^{+2.8}$ found with the Reichart method. 
Likewise, using the sample of 551 GRBs with pseudo-redshift of G05,
we find the best fit parameter values reported in Table~\ref{tab:fit_results},
$m=0.88_{-0.13}^{+0.12}$. 
The best-fitting power-laws obtained with the D'Agostini method
and the Reichart method are compared in Fig.s~\ref{f:z} and \ref{f:pz}
for the two data sets, respectively.

In Table~\ref{tab:fit_results} we report the
also the best-fitting parameters obtained using the
Least square fit in the case of data 
in the case of data affected by errors on both axes (``{\tt fitexy}'' tool, 
\citet{Press93}), but not with no extra variance. We report these results
to examine how the best-fitting parameters are affected
when the sample variance is taken into account.

As can be seen from Table~\ref{tab:fit_results}, the Reichart method 
and D'Agostini method give different best fit parameter values,
especially the value of the slope $m$.
The Reichart method yields steeper slopes even that those found
with the D'Agostini method,
whose results are consistent with those originally obtained by GFM05
and G05 ($m = 1.3_{-0.4}^{+0.8}$, GFM05; $m = 0.85\pm 0.02$, G05),
even though the fit, in the latter case, was found unacceptable 
(very high $\chi_r^2$).
Also with the '{\tt fitexy}' algorithm \citep{Press93}, the slopes 
obtained are much shallower than the correspondent obtained with the
Reichart method.

%%%%%%%%%%%%%%%%%%%%%%%%%%%%%%%%%%%%%%%%%%%%%%%%%%%%%%%%%
\section{Comparative Assessment of Analysis Methods}
\label{s:lim}
%%%%%%%%%%%%%%%%%%%%%%%%%%%%%%%%%%%%%%%%%%%%%%%%%%%%%%%%%
In order to test the consistency of the estimate of the $m$ parameter
obtained with each method, we performed various numerical simulations.

Assuming the existence of a linear correlation between
$y = \log{(L/\bar{L})}$ and $x = \log{(V/\bar{V})}$ with prefixed 
parameters ($m$, $q$, $\sigma_x$, $\sigma_y$), using random number
generators, proper
data sets were produced and the data analysis was performed following both
the Reichart and the D'Agostini methods.
The best-fit parameters were thus compared with
the corresponding input values, also called ``true values''.
We assumed $\sigma_y=0$ without loss of generality, as discussed in
Sec.~\ref{s:appl}. We call $\sigma_{x,{\rm true}}$ the value assumed
to generate the populations of random $(x_i,y_i)$ points, in order
to distinguish it from the best-fitting values derived with each method.
Likewise, we denote $m_{\rm true}$ the correct value for $m$.

From our tests it resulted that there are two different regimes, mainly
depending on the value of the ratio between
the extrinsic variance $\sigma_x^2$
and the total variance of the sample along $x$, $\sigma_{x,t}^2$.

Concerning the distribution of intrinsic uncertainties along $x$ and $y$,
$\sigma_{x,i}$ and $\sigma_{y,i}$ ($i=1,\ldots,N$), respectively,
we considered two different cases:
\begin{enumerate}
\item the \{$\sigma_{x,i}$\} have been generated from the
distribution of the intrinsic uncertainties along $x$ of the sample
of 32 GRBs of GFM05. Figure~\ref{f:sigma_xi_vs_x} shows the two
populations of $\sigma_{x,i}$ as a function of $x$ in two cases:
the real case of the 32 GRBs of GFM05 and a random population of
200 points. The same was done along $y$ to generate the distribution
of $\sigma_{y,i}$; the result is shown in Fig.~\ref{f:sigma_yi_vs_y}.
%
%Figure 3
%
\begin{figure}
\includegraphics[width=8.5cm]{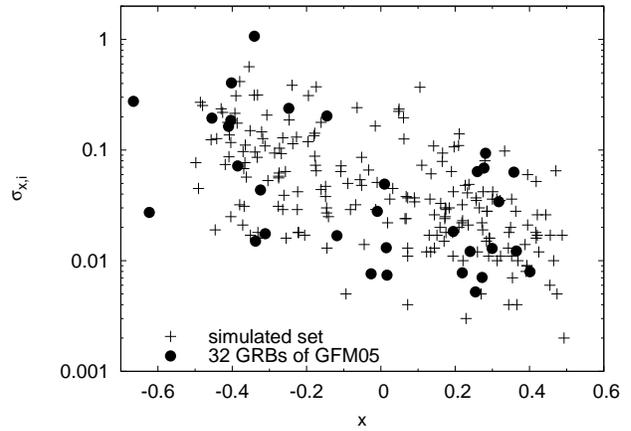}
\caption{Intrinsic uncertainties $\sigma_{x,i}$ as a function
of $x$ for two populations: 200 random simulated points (crosses)
and the 32 real GRBs of GFM05 (filled circles).}
\label{f:sigma_xi_vs_x}
\end{figure}
%
%Figure 4
%
\begin{figure}
\includegraphics[width=8.5cm]{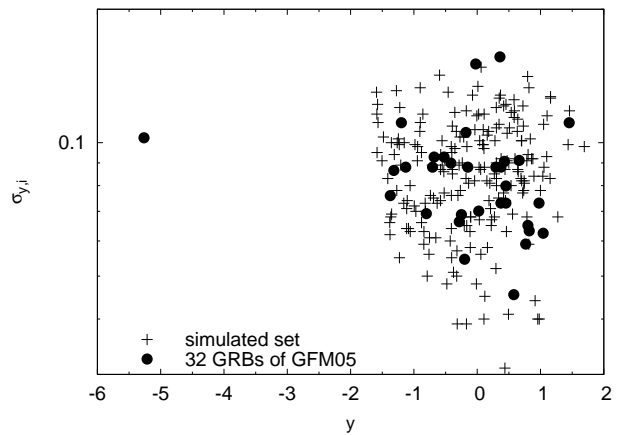}
\caption{Intrinsic uncertainties $\sigma_{y,i}$ as a function
of $y$ for two populations: 200 random simulated points (crosses)
and the 32 real GRBs of GFM05 (filled circles).}
\label{f:sigma_yi_vs_y}
\end{figure}
\item the same as the previous case, but with a population of
intrinsic uncertainties \{$\sigma_{x,i}$\} a factor $1/10$ times
as big as those generated following the previous method.
The reason is to study the regime
$\overline{\sigma_{x,i}^2}<<\sigma^2_{x,t}$.
\end{enumerate}

For each simulated data set, we generated 1000 points according
to a uniform distribution in $x$ in the range $-2<x<2$.
We then selected 200 points in the range $-0.5<x<0.5$.
This choice is because
we want a range along $x$ of 1, which is the same as that of the
32 GRBs of GFM05, ranging in $-1.6<\log{V}<-0.6$ (Fig.~\ref{f:z}).

Concerning $\sigma_{x,{\rm true}}$, we studied the dependence
of $m$ and $\sigma_{x}$ as a function of $\sigma_{x,{\rm true}}$
in the range $0<\sigma_{x,{\rm true}}<0.5$.

\begin{itemize}
\item{{\bf Case a}}.
We assumed $m_{\rm true}=1.7$, $q_{\rm true}=0$
(similar to that derived with the D'Agostini method for the GFM05 sample),
and a distribution of intrinsic uncertainties along $x$ and $y$, \{$\sigma_{x,i}$\}
and \{$\sigma_{y,i}$\}, given by item (i) (Figs.~\ref{f:sigma_xi_vs_x}
and \ref{f:sigma_yi_vs_y}). The estimated $m$ as a function of the
true extrinsic scatter along $x$, $\sigma_{x,{\rm true}}$, is shown
in Fig.~\ref{f:m1.7_m_vs_ratiosigma}, for either method.
%
%Figure 5
%
\begin{figure}
\includegraphics[width=8.5cm]{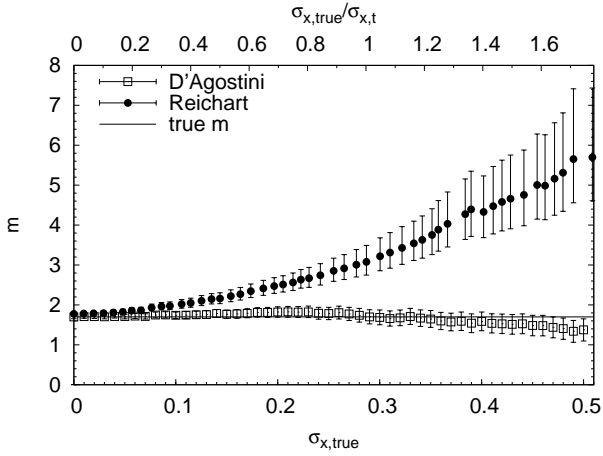}
\caption{$m$ as a function of $\sigma_{x,{\rm true}}$. Solid
line shows $m_{\rm true}=1.7$ ({\bf Case a}).
Empty squares correspond to the
D'Agostini, filled circles to the Reichart method. Error bars
show the 90\% confidence interval.}
\label{f:m1.7_m_vs_ratiosigma}
\end{figure}
Apparently, when the ratio $\sigma_{x,{\rm true}}/\sigma_{x,t}>0.4$,
$m$ estimated with the Reichart method is significantly overestimated,
while that derived with the D'Agostini method turns out to be a more
consistent estimator.

Figure~\ref{f:m1.7_sigmatrue_vs_sigmaest} shows $\sigma_x$ derived
with each method as a function of the true value $\sigma_{x,{\rm true}}$.
% Figure 6
\begin{figure}
\includegraphics[width=8.5cm]{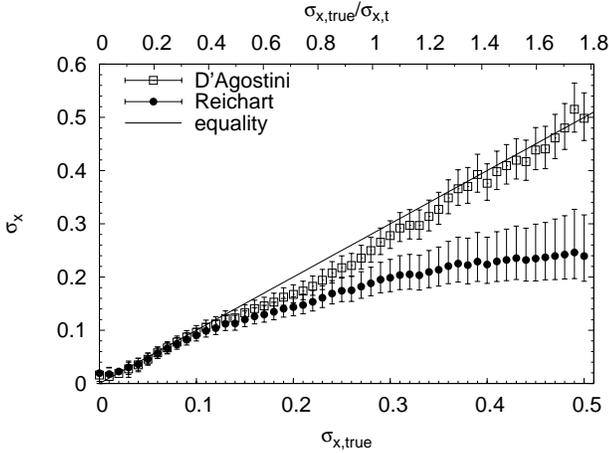}
\caption{$\sigma_x$ estimated with either method as a function
of the true value $\sigma_{x,{\rm true}}$ ({\bf Case a}).}
\label{f:m1.7_sigmatrue_vs_sigmaest}
\end{figure}
In this example, for $\sigma_{x,{\rm true}}/\sigma_{x,t}>0.4$,
the Reichart method seems to underestimate the
extrinsic scatter $\sigma_{x}$ (Fig.~\ref{f:m1.7_sigmatrue_vs_sigmaest})
and overestimate the slope $m$ (Fig.~\ref{f:m1.7_m_vs_ratiosigma});
these two aspects appear to be correlated.
Figure~\ref{f:m1.7_sigx0.34_x_vs_y} shows the case of $\sigma_{x,{\rm true}}=0.34$,
similarly to the result of the D'Agostini method applied to the GFM05
sample. In that case, the estimated slopes result: $m=3.6_{-0.4}^{+0.6}$
(Reichart), $m=1.7\pm0.2$ (D'Agostini). The total scatter
along $x$ is $\sigma_{x,t}=0.285$.
%Figure 7
%
\begin{figure}
\includegraphics[width=8.5cm]{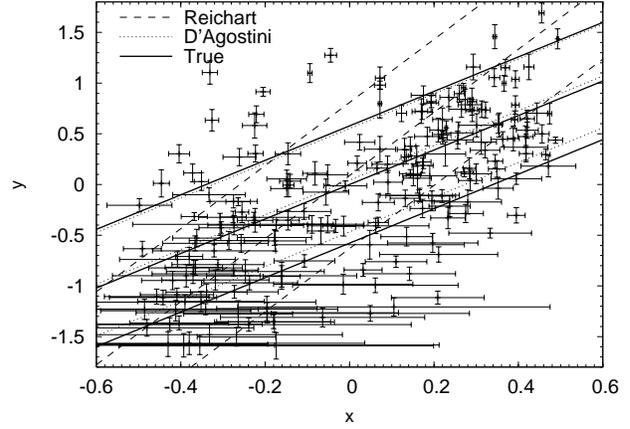}
\caption{Sample of 200 simulated points generated assuming
$m_{\rm true}=1.7$, $q_{\rm true}=0$ ({\bf Case a}) and $\sigma_{x,{\rm true}}=0.34$.
Dashed lines show the best-fit and 1$\sigma$ region obtained
with the Reichart, while dotted lines correspond to the D'Agostini method.
Solid thick lines show the true power law and dispersion.}
\label{f:m1.7_sigx0.34_x_vs_y}
\end{figure}

\item{{\bf Case b}}. 
We adopted the same values for $m_{\rm true}$ and $q_{\rm true}$
as those of {\bf Case a}, but with intrinsic errors 10 times smaller
along both axes, as described by (ii).
The results are basically the same as {\bf Case a},
confirming that when $\sigma_{x,{\rm true}}$ becomes comparable
with $\sigma_{x,t}$ the Reichart method overestimates $m$
and underestimates $\sigma_{x}$. When $\sigma_{x,{\rm true}}=0.34$
it results: $m=3.7\pm0.5$ and $\sigma_x\sim0.22$ (Reichart),
$m=1.6\pm0.2$ and $\sigma_x\sim0.35$ (D'Agostini).

\item{{\bf Case c}}. 
We assumed $m_{\rm true}=3.8$, $q_{\rm true}=0$
(similar to that derived with the Reichart method for the GFM05 sample),
and a distribution of intrinsic uncertainties along $x$ and $y$, \{$\sigma_{x,i}$\}
and \{$\sigma_{y,i}$\}, given by item (i) (Figs.~\ref{f:sigma_xi_vs_x}
and \ref{f:sigma_yi_vs_y}). The estimated $m$ as a function of the
true extrinsic scatter along $x$, $\sigma_{x,{\rm true}}$, is shown
in Fig.~\ref{f:m3.8_m_vs_ratiosigma}, for either method.
%
%Figure 8
%
\begin{figure}
\includegraphics[width=8.5cm]{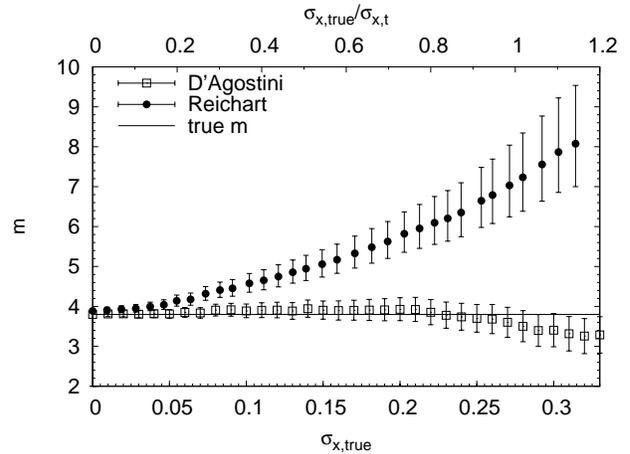}
\caption{$m$ as a function of $\sigma_{x,{\rm true}}$. Solid
line shows $m_{\rm true}=3.8$ ({\bf Case c}).
Empty squares correspond to the
D'Agostini, filled circles to the Reichart method. Error bars
show the 90\% confidence interval.}
\label{f:m3.8_m_vs_ratiosigma}
\end{figure}
Similarly to what found in {\bf Case a},
when the ratio $\sigma_{x,{\rm true}}/\sigma_{x,t}>0.2$,
$m$ estimated with the Reichart method is significantly overestimated,
while that derived with the D'Agostini method proves to be a more robust
estimator.

Figure~\ref{f:m3.8_sigx0.24_x_vs_y} shows the case of $\sigma_{x,{\rm true}}=0.24$,
similarly to the result of the Reichart method applied to the GFM05
sample. In that case, the estimated slopes result: $m=6.2_{-0.6}^{+0.7}$
and $\sigma_x\sim0.17$ (Reichart),
$m=3.7_{-0.3}^{+0.4}$ and $\sigma_x\sim0.22$ (D'Agostini).
The total scatter along $x$ is $\sigma_{x,t}=0.286$.
%Figure 9
%
\begin{figure}
\includegraphics[width=8.5cm]{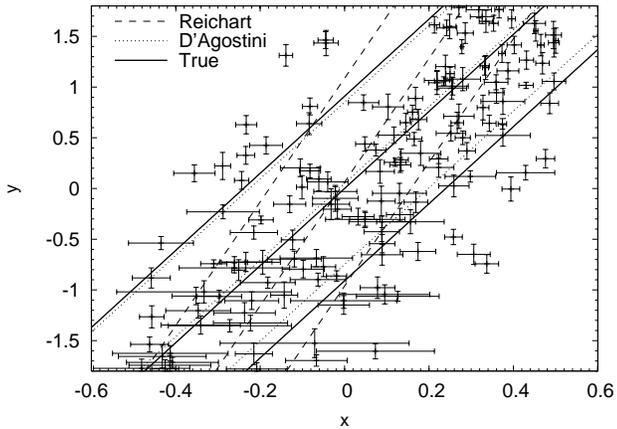}
\caption{Sample of 200 simulated points generated assuming
$m_{\rm true}=3.8$, $q_{\rm true}=0$ ({\bf Case c}) and $\sigma_{x}=0.24$.
Dashed lines show the best-fit and 1$\sigma$ region obtained
with the Reichart, while dotted lines correspond to the D'Agostini method.
Solid thick lines show the true power law and dispersion.}
\label{f:m3.8_sigx0.24_x_vs_y}
\end{figure}

\item{{\bf Case d}}. 
We adopted the same values for $m_{\rm true}$ and $q_{\rm true}$
as those of {\bf Case c}, but with intrinsic errors 10 times smaller
along both axes, as described by (ii).
The results are consistent with those of {\bf Case c},
confirming that when $\sigma_{x,{\rm true}}$ becomes comparable
with $\sigma_{x,t}$ the Reichart methods overestimates $m$
and underestimates $\sigma_{x}$. When $\sigma_{x,{\rm true}}=0.24$
it results: $m=6.4\pm0.6$ and $\sigma_x\sim0.18$(Reichart),
$m=3.9_{-0.3}^{+0.4}$ and $\sigma_x\sim0.22$ (D'Agostini).
\end {itemize}

\begin{figure*}
\includegraphics[width=17cm]{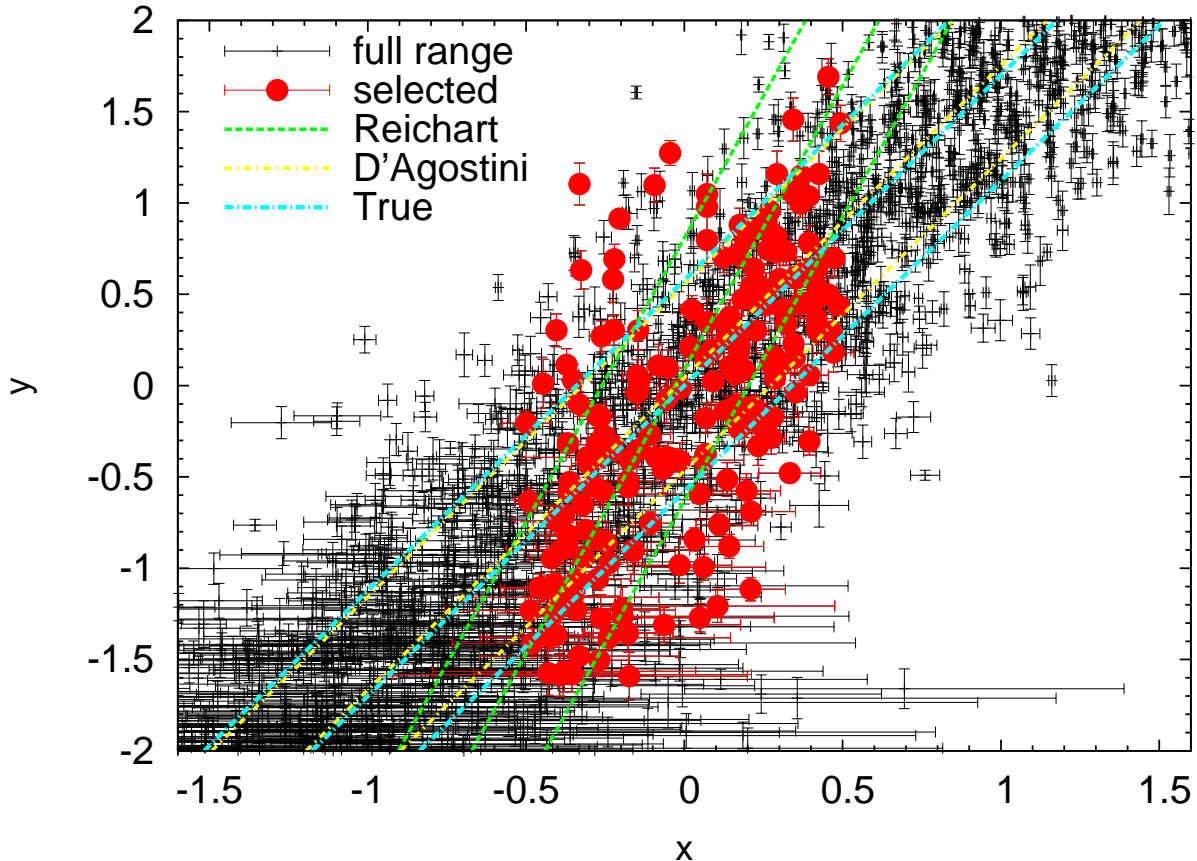}
\caption{Full population of 1000 simulated points (crosses)
from {\bf Case a}, with $m_{\rm true}=1.7$ and $\sigma_{x,{\rm true}}=0.34$.
Filled red circles show the selected subsample of 200 points 
$-0.5<x<0.5$, both fitting methods have been applied to in
{\bf Case a}.}
\label{f:both_m1.7_0.34}
\end{figure*}

To better appreciate what happens when the total scatter
$\sigma_{x,t}$ of the sample becomes comparable with the
extrinsic scatter $\sigma_{x}$, as it might be the case when
truncation effects are at work, in Fig.~\ref{f:both_m1.7_0.34} we show
the the entire population of 1000 random points generated
in {\bf Case a}, when $\sigma_{x,true}=0.34$, in the range
$-2<x<2$ and the selected subsample of 200 points in the range
$-0.5<x<0.5$. The total scatter $\sigma_{x,t}$ is 1.39
($\sigma_{x,true}/\sigma_{x,t}=0.24$) for the entire sample
of 1000 points and is 0.285 ($\sigma_{x,true}/\sigma_{x,t}=0.84$)
for the subsample of 200 points.
%Figure 10
%
%

%%%%%%%%%%%%%%%%%%%%%%%%%%%%%%%%%%%%%%%%%%%%%%%%%%%%%%%%%
\section{Discussion}
%%%%%%%%%%%%%%%%%%%%%%%%%%%%%%%%%%%%%%%%%%%%%%%%%%%%%%%%%
\label{s:disc}

The results obtained from the numerical simulations show that,
when the sample variance $\sigma_{x,{\rm true}}^2$ is significantly lower than 
the total variance $\sigma_{x,t}^2$, both methods provide
a consistent estimate of the $m$ slope.
Differently, when $\sigma_{x,{\rm true}}^2$ is comparable with $\sigma_{x,t}^2$,
the Reichart method overestimates $m$, while
the D'Agostini method still gives an estimate of $m$ compatible with
the correct value $m_{\rm true}$.

The simulations have been performed adopting the same distributions
of intrinsic errors along both axes, \{$\sigma_{x,i}$\} and
\{$\sigma_{y,i}$\} $(i=1,\ldots,N)$, as those of the sample of 32
GRBs of GFM05. For each simulated data set, we selected 200 points
within the range $-0.5<x<0.5$, similarly to the range covered in
$\log{(V/\bar{V})}$ by the 32 GRBs of GFM05.
Typical values of the total scatter $\sigma_{x,t}$ for the simulated
data sets are 0.28--0.29, to be compared with that of the 32 GRBs
of GFM05, which is 0.32.
The results obtained are confirmed even when we adopt the same
extrinsic scatter and 10 times smaller intrinsic errors along
either axis. The most important factor is the ratio between
the extrinsic scatter, $\sigma_{x,{\rm true}}$, and the
total scatter, $\sigma_{x,t}$.

The two methods differ when the range along $x$ is comparable
with the extrinsic scatter along the same axis.
This is shown in Fig.~\ref{f:both_m1.7_0.34}: the full
population clearly follows the true slope of $m_{\rm true}=1.7$
shown by the cyan line. However the two methods, applied to the
200 points selected in the range $-0.5<x<0.5$, for which the ratio
$\sigma_{x,{\rm true}}/\sigma_{x,t}=0.84$, give different results,
the Reichart method (green) overestimating $m$, while
the D'Agostini method (yellow) is still consistent with the true
slope.
We do not know yet whether the $V-L$ correlation presently known
extends along $V$ over a wider range,
the samples considered by \citet{Reichart01}
and GFM05 being the result of truncation effects, or it does not.
Either way, however, the total variance is comparable
with the extrinsic variance of the correlation and in this case, the
D'Agostini method turns out to be a more reliable estimator of $m$
than the Reichart.

Similar results to those presented here and by GFM05 and G05,
i.e. values of $m$ around 1--2, have already been mentioned by
\citet{Lloyd02a}: in fact, therein we read that a more recent
analysis by \citet{Fenimore00} making use of a different definition
of variability, based on a different degree of smoothing of the
light curve, led to an estimate of $m=1.57\pm\sim0.5$.
This supports the view of a not well established value of power-law
index of the $V-L$ correlation, affected by a significant scatter
\citep{Lloyd02a}.

In another recent paper by \citet{Li06} a new definition of
variability is considered, based on a different smoothing filter
of the light curves. In this work, they find a slope of $\sim$3,
applying the {\tt fitexy} procedure, and no extra variance is needed,
as the correlation appears to be tight enough.
These results as well as the recent analysis by \citet{Fenimore00}
point to a strong dependence of $m$ on the kind of filter adopted
to obtain a smoothed version of the light curve,
with respect to which variability of the original light curve
is computed.

As far as the definition of variability introduced by \citet{Reichart01}
is regarded, we have investigated reasons for the overestimation of $m$
given by the Reichart method and found that this
is likely an incorrect derivation of the likelihood function 
by Reichart (2001) (eq.~\ref{e:reichart}). This function was derived by
integrating the weighted product of Gaussian density functions over, 
among other variables, the element $ds^2 = (dx^2 + dy^2)$ (eq. 28 of Reichart 2001),
where $y = y_c(x, \theta_m)$ is the equation which describes the relation
between the parameters $x$ and $y$ (see Section~\ref{s:meth}).
The integration over $ds^2 = (dx^2 + dy^2)$ is not correct: $dy$ is
homogeneous with $m\,dx$ and not with just $dx$.
The integration on $ds$ introduces the sum $1+m^2$ in the numerator of
the first term at right hand side of eq.~\ref{eq:prior}, which is the
only difference between the likelihood functions of the two methods.
The inconsistency of the term $ds$ mentioned above
is transmitted to the factor $1+m^2$: $m$ cannot be added to 1 tout court
for the same reason explained above and discussed by \citet{Dagostini05}.
The effect of this factor is the overestimation of $m$.
This is shown by the fact that,
by deleting $m^2$, the estimate of $m$ becomes consistent and
 the resulting likelihood function becomes coincident with that
derived by \citet{Dagostini05} for the same statistical inference problem,
reported here in eq.~\ref{eq:prior_dago}.
 
The procedure adopted by GFM05 and G05 as well as the usage of the
{\tt fitexy} procedure reported in Table~\ref{tab:fit_results}
does not take into account the sample or extrinsic variance.
As a result they find a high $\chi^2$ 
assuming a straight line for the correlation between $\log L$ 
and $\log V$. However, if, in spite of that, a straight line is assumed,
the line slope $m$ value found is very close to the true one.

We conclude that the results reported by R01 and RN05, who applied
the Reichart method, give an overestimated index $m$ of the power law ($L\propto V^m$)
function used to describe the correlation between peak luminosity
and variability of the GRBs included in the sample of R01, GFM05 and
G05. We ascribe the different results obtained with the D'Agostini
and Reichart methods, when applied to the samples of 32 GRBs of GFM05
and 551 GRBs of G05, to the comparable total variance $\sigma_{x,t}^2$
and extrinsic variance $\sigma_x^2$.
The D'Agostini method appears to provide the most correct
estimations of the slope and extrinsic scatter of either sample.
The best-fitting power laws derived by GFM05 and G05 are more
consistent with the results of the D'Agostini than with the Reichart
method.

The implications of different power-law index and extrinsic
scatter describing the $V-L$ correlation are twofold:
the physical interpretation and its usage as a luminosity estimator.

%%%%%%%%%%%%%%%%%%%%%%%%%%%%%%%%%%%%%%%%%%%%%%%%%%%%%%%%%
\subsection{Implications on the interpretation}
\label{s:impl_int}
%%%%%%%%%%%%%%%%%%%%%%%%%%%%%%%%%%%%%%%%%%%%%%%%%%%%%%%%%
Variability in the GRB light curves is thought to be produced
mainly above the photospheric radius at which the relativistic
outflow becomes optically thin to electron scattering.
In addition to this photosphere, there would be a second photospheric
radius beyond which shocks are optically thin to pairs too
and highly variable light curves originate.
\citet{Meszaros02} interpreted the $V-L$ correlation as due
to the effects of this pair-forming photosphere and the
jet opening and/or viewing angles, when breaks are observed
in the optical consistent with the jet interpretation.
The power-law $m$ derived by those authors is $m=6p/(5q-2p)$,
where they assumed $L\propto\theta^{-p}$, $p\sim 2-2.5$
($\theta$ can be either the jet opening angle or the viewing angle
for a universal jet pattern), $\Gamma_m\propto\theta^{-q}$
($\Gamma_m$ is the minimum value of the Lorentz factor distribution
of the shells).
Following \citet{Meszaros02}, if one takes $p=q$, the expected
power-law index is $m=2$, which is consistent within 90\% CL
with the value we derived with the D'Agostini method.
\citet{Meszaros02} assumed $p=5/2$ and $q=2$ to explain the
value originally found by \citet{Reichart01} and by
\citet{Fenimore00} of $m\sim 3$.
\citet{Kobayashi02} started from similar assumptions:
$\Gamma\propto\theta^{-q}$, i.e. a wider jet has a lower Lorentz
factor, equal mass colliding shells, a uniform $D/c$ distribution
in logarithmic spaces in the interval 1~ms--1~s ($D$ is
the separation between shells). Using a simplified version
of the \citet{Reichart01} variability they tried to reproduce the $V-L$
correlation with numerical simulations and found that $q=1$
seems to better account for $m$ values in the range of 3--4,
as found by \citet{Reichart01} and \citet{Fenimore00}.
However, from the same numerical results it turns out that
smaller values of $m$, compatible with the results obtained
with the D'Agostini method, are given by assuming $q=2$.
As already pointed out by \citet{Kobayashi02}, this is more consistent
with the value $q=8/3$ derived by \citet{Salmonson02}, based
on the study of the anticorrelation between peak luminosity and
jet break time. From the results of \citet{Kobayashi02}, 
we infer that the results of the D'Agostini best-fitting parameter
$m$ imply a stronger dependence of the Lorentz factor
on the opening angle as well as a smaller Lorentz factor normalisation.

Likewise, \citet{Ramirez02} studied the effects of varying the energy
per unit solid angle in the collimated outflow on the correlations between
variability, peak luminosity and spectral peak energy \citep{Lloyd02b}.
They found that these correlations are accounted for if luminosity per
unit solid angle $L(\theta)$ (where $\theta$ can be either the viewing
angle for a structured jet pattern or the jet opening angle),
strongly depends on the Lorentz factor $\Gamma(\theta)$.
while the effect of changes in the baryon loading of the wind is less
relevant, as it does not account for the range of variability observed.
In their figure~3, \citet{Ramirez02} show the results in the $V-L$ space
of some numerical simulations for different bulk Lorentz factors:
values for $m$ in the range 1.5--2, i.e. like those found by GFM05 and
those derived in this work with the D'Agostini method, are clearly
compatible with the results of these simulations.

The smaller value of $m$ we found with the D'Agostini method
than with the Reichart constrains differently the parameters
$m$ depends on. Moreover, the bigger extrinsic
scatter of $\sigma_{\log{V}}$, 0.34 with the D'Agostini,
0.24 with the Reichart method, means a bigger scatter in
the distribution of some of these parameters.

%%%%%%%%%%%%%%%%%%%%%%%%%%%%%%%%%%%%%%%%%%%%%%%%%%%%%%%%%
\subsection{Implications on its usage as luminosity estimator}
\label{s:impl_candle}
%%%%%%%%%%%%%%%%%%%%%%%%%%%%%%%%%%%%%%%%%%%%%%%%%%%%%%%%%
After the discovery of the $V-L$ correlation by \citet{Fenimore00}
and \citet{Reichart01}, a number of papers used it
as a luminosity estimator for several studies.
\citet{Fenimore00} applied it to derive the luminosity distribution
$\phi(L)$ and the GRB formation rate as a function of redshift
$\rho(z)$. \citet{Lloyd02a} studied the bivariate distribution
of luminosity and redshift $\Psi(L,z)$ using the pseudo-redshifts
obtained from a parametrisation of the $V-L$ correlation
with $m=1.57\pm\sim0.5$.
They found a correlation between luminosity and redshift
parametrised as $L\propto (1+z)^{1.7\pm0.4}$.
Different values of $m$ from $\sim1$ to $\sim3$
do not change their results qualitatively \citep{Lloyd02a}.

\citet{Lazzati02} exploited the $V-L$ correlation
to study the rest-frame power spectral density (PSD) in time domain
for hundreds of BATSE bursts. In particular, the goal was to
study the effect of photon scattering on variability.
A correlation between the break frequency in the rest-frame PSD 
and variability was found, based on $m\sim3$.
A smaller value of $m$ would mean that a GRB with a given
variability, whose dependence on $z$ is small, would be
less luminous, i.e. nearer, thus with a corresponding smaller
break frequency in the rest-frame PSD.
This relation would have to be rescaled consequently.

More recently, other tighter correlations became popular,
such as between the rest-frame peak energy $E_{\rm p}$ and
the isotropic-equivalent gamma-ray released energy
$E_{\gamma,{\rm iso}}$,
$E_{\rm p}\propto E_{\gamma,{\rm iso}}^{0.5}$ \citep{Amati02},
between $E_{\rm p}$ and the collimation-corrected gamma-ray
released energy $E_\gamma$, 
$E_{\rm p}\propto E_\gamma^{0.7}$ \citep{Ghirlanda04}.
Another relation between $E_{\rm p}$, $E_{\gamma,{\rm iso}}$
and the rest-frame break time $t_{\rm b}$ of the optical
afterglow light curve has been discovered recently
\citep{Liang05}, which requires fewer assumptions than
those of the Ghirlanda relationship.
The same relation as the Amati holds with the same index
of $\sim$0.5 when $E_{\gamma,{\rm iso}}$
is replaced by the peak luminosity $L_{\gamma,{\rm iso}}$, or
simply $L$ \citep{Yonetoku04,Ghirlanda05}.
Recently, the Amati relation has been confirmed for a sample
of 53 long GRBs and a fit with the D'Agostini method gives
an extrinsic logarithmic scatter of 0.15 over 3 orders of
magnitude over $E_{\rm p}$ \citep{Amati06}.
In this case both methods give compatible results and this
is consistent with the fact that $\sigma_{x}<<\sigma_{x,t}$.
Summing up, in the case of the Amati relation,
truncation effects are negligible.

Interestingly, \citet{Lloyd02b} studied the correlation
between $V$ and $E_{\rm p}$ before the discovery of the
Amati relationship. Assuming different alternative values
for $m$, ranging from 2.2 to 5.8, as originally found
by \citet{Fenimore00}, they found a correlation,
$E_{\rm p}\propto V^{\delta}$, with $\delta$ ranging
from 0.4 to 1.15 . The Amati relation,
whose extrinsic scatter is small with respect to that of
the $V-L$ correlation, is more reliable as a luminosity
estimator. Hence, from the Amati relation we should
expect $m=\delta/0.5$, so in the range 0.8--2.3, which
is consistent with the results of the D'Agostini
method as well as those of GFM05 and G05.

%%%%%%%%%%%%%%%%%%%%%%%%%%%%%%%%%%%%%%%%%%%%%%%%%%%%%%%%%
\section{Conclusions}
\label{s:conc}
%%%%%%%%%%%%%%%%%%%%%%%%%%%%%%%%%%%%%%%%%%%%%%%%%%%%%%%%%
We applied both Reichart and D'Agostini methods to the samples
of 32 GRBs with known redshift and 551 BATSE GRBs with
pseudo-redshift considered by GFM05 and G05, respectively.
The goal was to estimate the slope $m$ as well as the scatter
of the power law describing the correlation between variability
and peak luminosity originally presented by \citet{Reichart01}.
Both methods account for an extra variance in addition to
the intrinsic affecting each single point.
From simulations, we found that when the sample variance $\sigma_x^2$
is comparable with the total scatter along the same axis, $\sigma_{x,t}^2$,
the Reichart method tends to overestimate $m$, while
the D'Agostini still estimates it correctly. When the sample
variance is negligible with respect to the total variance, the
two methods give consistent results.
In the specific case of the $V-L$ correlation, the two variances
are comparable: in the case of the 32 GRBs of GFM05, it is
$\sigma_{x,t}=0.32$, $\sigma_{x}=0.24$ (Reichart) or 
$\sigma_{x}=0.34$ (D'Agostini). This explains the discrepancy
between the two methods. We showed that the D'Agostini method
is a reliable estimator of $m$ in this regime, whereas the
Reichart is no more. In particular, the best-fitting value for
$m$ obtained with the D'Agostini method is $1.7\pm0.4$ and
$0.88_{-0.13}^{+0.12}$ at 90\% confidence level for the 32 GRBs
of GFM05 and for the 551 GRBs of G05, respectively.
These values are significantly smaller than those obtained
with the Reichart method, which are consistent with previous results
of R01 and RN05.
These results hold as far as the definition of variability given by
R01 is assumed. 

Alternatively, from other definitions of variability based on different
kinds of filter used to smooth the light curves, it seems to be
possible to obtain a range of values for $m$ from $\sim$1 to $\sim$3
\citep{Fenimore00,Li06}.

One of the possible implications of a smaller value of $m$ than
originally found by R01 on the interpretation of the $V-L$ correlation
is that, in the jet emission scenario, we would expect a stronger
dependence of the Lorentz factor of the expanding shells on the
jet opening angle \citep{Kobayashi02}.

Finally, other more recent and tighter correlations, such as the
Amati, Ghirlanda, Liang \& Zhang relations appear to be more reliable
luminosity estimators than the $V-L$ one. In particular, our
results of values of $m$ in the interval 1--2 obtained with the
D'Agostini method, appear to be consistent with two independent
relations: $E_{\rm p}\propto V^{\delta}$ ($0.4<\delta<1.15$)
\citep{Lloyd02b},
and the equivalent version of the Amati relation with the
isotropic-equivalent peak luminosity instead of the
isotropic-equivalent total released energy,
$E_{\rm p}\propto L^{0.5}$. By combining the two, one expects
$m=\delta/0.5$, in agreement with the results reported here
as well as those reported by GFM05 and G05.

Finally, the increasing number of GRBs with spectroscopic
redshift detected by {\em Swift} \citep{Gehrels04}
will help extend the range
in $V$ and better constrain the power-law fit of the $V-L$
correlation, with the benefit of a homogeneous data set
of light curves all detected with the Burst Alert Telescope
(BAT) onboard {\em Swift}. A thorough test of the $V-L$ correlation
with BAT data is in progress (Rizzuto et al., in preparation).

%
% two kinds of implications:
% interpretation of the V-L correlation (Kobayashi, Meszaros)
% its usage as a luminosity estimator
%
%

%%%%%%%%%%%%%%%%%%%%%%%%%%%%%%%%%%%%%%%%%%%%%%%%%%%%%%%%%
\section*{Acknowledgments}
%%%%%%%%%%%%%%%%%%%%%%%%%%%%%%%%%%%%%%%%%%%%%%%%%%%%%%%%%

We thank the anonymous referee for useful comments.
C.G. thanks S. Kobayashi for useful discussions.
C.G. acknowledges his Marie Curie Fellowship from the European Commission.
C.G., F.F., E.M., F.R. and L.A. acknowledge support from
the Italian Space Agency and Ministry of University and
Scientific Research of Italy (PRIN 2003 on GRBs).
A.G. acknowledges the receipt of Marie Curie European Re-integration Grant.
C.G.M. acknowledges financial support from the Royal Society.
This research has also made use of data obtained from the {\it HETE2} science team via
the website http://space.mit.edu/HETE/Bursts/Data and BATSE, Konus/{\it WIND} and
BAT/{\it Swift} data obtained from the High-Energy Astrophysics Science Archive
Research Center (HEASARC), provided by NASA Goddard Space Flight Center.

%%%%%%%%%%%%%%%%%%%%%%%%%%%%%%%%%%%%%%%%%%%%%%%%
% BIBLIOGRAPHY
%%%%%%%%%%%%%%%%%%%%%%%%%%%%%%%%%%%%%%%%%%%%%%%%

\bsp

\label{lastpage}

\end{document}